\documentstyle[12pt,epsf,epsfig]{article}
\def\beq{\begin{equation}}
\def\eeq{\end{equation}}
\def\bea{\begin{eqnarray}}
\def\eea{\end{eqnarray}}
\def\bq{\begin{quote}}
\def\eq{\end{quote}}

\def\APP{{\it Acta Phys.Pol.} }

\def\NP{{\it Nucl.Phys.} }
\def\PL{{\it Phys.Lett.} }
\def\PR{{\it Phys.Rev.} }

\def\SJNP{{\it Soviet J.Nucl.Phys.} }

\def\SP{{\it Soviet.Phys.} }

\def\YF{{\it Yadernaya Fizika} }

\def\ZP{{\it Z.Phys.} }

\def\gappeq{\mathrel{\rlap {\raise.5ex\hbox{$>$}}
{\lower.5ex\hbox{$\sim$}}}}

\def\lappeq{\mathrel{\rlap{\raise.5ex\hbox{$<$}}
{\lower.5ex\hbox{$\sim$}}}}

\begin{document}
\pagestyle{empty}
\begin{flushright}
{CERN-TH.95-237}
\end{flushright}
\vspace*{5mm}
\begin{center}
{\bf $\mbox{\boldmath$k$}$-FACTORIZATION}$^{*)}$ \\
\vspace*{1cm}
{\bf Marcello Ciafaloni}$^{+)}$ \\
\vspace{0.3cm}
Theoretical Physics Division, CERN \\
CH - 1211 Geneva 23 \\
\vspace*{2cm}
{\bf ABSTRACT} \\ \end{center}
\vspace*{5mm}
\noindent
I review the $\mbox{$\boldmath k$}$-factorization method to combine
the
high-energy behaviour in QCD with the
renormalization group. Resummation formulas for coefficient functions
and anomalous dimensions
are derived, and their applications to small-$x$ scaling violations
in
structure functions are
briefly discussed.

\vspace*{5cm}

\noindent

\rule[.1in]{16.5cm}{.002in}

\noindent
$^{*)}$ Talk delivered at the VIth Blois Workshop ``Frontiers in
Strong
Interactions",
June 20-24, 1995.

\noindent
$^{+)}$ On sabbatical leave of absence from Dipartimento di Fisica,
Universit\`a
di Firenze, and INFN, Sezione di Firenze, Italy.
\vspace*{0.5cm}

\begin{flushleft} CERN-TH.95-237 \\
August 1995
\end{flushleft}
\vfill\eject

high-energy behaviour in QCD with the
and anomalous dimensions
structure functions are


\setcounter{page}{1}
\pagestyle{plain}

\section{Introduction}

Small-$x$ hard processes are characterized by a large scale $Q^2 \gg
\Lambda^2$,
and by a much larger energy $s = Q^2/x \gg Q^2$. Therefore,
renormalization
group (RG) factorization and high-energy,
$\mbox{\boldmath$k$}$-dependent,
factorization should both apply to deep inelastic scattering (DIS) in
the small
Bjorken $x$ region explored, e.g., at HERA$^{1)}$. This talk is about
their
consistency and its consequences for QCD perturbative predictions.

The approach of $\mbox{\boldmath$k$}$-factorization$^{2)-6)}$,
developed
by
Camici, Catani, Hautmann and myself, and triggered by a work of
Nason,
Dawson
and Ellis$^{7)}$, leads to resummation formulas of QCD perturbative
results for
coefficient functions$^{2),3)}$ and for anomalous dimensions$^{4)}$.
It
also
leads to a generalization of the effective $W$-approximation in
electroweak
fusion processes$^{5)}$. Here I will briefly review its basic
results,
by adding
a few comments$^{6)}$ on their use to explain the HERA data$^{8)}$.

The high-energy approach and the RG one have a long
history$^{9)-11)}$
and are,
to start with, largely different.

The former is valid for $s \gg Q^2$ and hadronic or partonic masses,
and
is
based on a diagrammatic analysis$^{9),10)}$ of gauge boson exchange
processes,
leading to quasi-constant high-energy cross-sections, of type
$(\alpha_s
\log
1/x)^n~~f_n~~(Q^2, Q^2_0)$, where $Q^2(Q^2_0)$ denotes the probe
(external
parton) virtuality.

The main outcome of this approach is the resummation of all leading
$\log x$ (LL) terms, provided
by the BFKL equation$^{9)}$ which predicts increasing
cross-sections$^{10)}$, due to a singularity
in the $t$-channel angular momentum $J > 1$ (the ``perturbative
Pomeron"), located at
\beq
J - 1 = \omega_P = \left({12\over\pi} \log 2\right) \alpha_s~.
\label{1}
\eeq
It is natural to think$^{12)}$ that such increase is related to the
small-$x$
rise of structure functions at HERA.

On the other hand, the RG approach is valid for $s, Q^2 \gg
\Lambda^2$
and $x =
Q^2/s$ fixed, and is based on the structure of collinear
singularities
for $Q^2
\gg Q^2_0$. It resums all terms $\alpha^n_s (\log Q^2/Q^2_0)^m
{}~~g_{nm}(x)$ for
$n \geq m > 0$, and it leads to the generalized GLAP
equations$^{11)}$,
which
yield the $\log Q^2$-evolution in terms of the QCD anomalous
dimensions
$\gamma_N(\alpha_s)$, where $N$ is the Mellin transform index in the
$x$
variable.

For single hard scale processes and high energies $(s \gg Q^2 \gg
Q^2_0)$, the
two methods above have to merge, with identification of $J$ and $N$
at
leading
$s$ level. On one hand, the BFKL equation develops collinear
singularities,
related to the $\log (Q^2/Q^2_0)$'s which have to be factorized out
and,
on the
other hand, the anomalous dimensions (and coefficient functions)
develop
$\alpha_s/(N-1)$ singularities, related to the $\log x$'s, which have
to
be
resummed. Since we are dealing with the same perturbative series, the
consistency requirement provides powerful constraints and new
results.

For instance, it is known that DIS structure functions, related to
total
cross-sections, have no $\alpha_s(\log x)^2$ terms in their
perturbative
expansion. This is almost trivial from the high-energy point of view:
there is
at most one emitted gluon per power of $\alpha_s$, and thus at most
one
power of
$\log (1/x)$ from rapidity integration. Instead, from the RG angle,
this
fact
requires refined cancellations$^{13)}$, and eventually implies that,
unlike the
timelike case$^{14)}$, the spacelike anomalous dimension $\gamma_N$
is a
function of the effective variable $\alpha_s/(N-1)$.

It is interesting to notice that such double $\log x$ terms are
instead
present in angular
distributions associated to DIS due to a new kind of form
factor$^{15),16)}$ and of small-$x$
equation$^{15)}$. They are of phenomenological interst for DIS event
generators based on small-$x$
branching schemes$^{17)}$ and call for quantitative studies$^{18)}$
in
the HERA energy range.

Limiting myself to total cross-sections, I will now summarize the
resummation formulas due to
$\mbox{\boldmath$k$}$-factorization, for both leading
$(\alpha_s/N-1)^n$
terms, and for some
next-to-leading (NL) ones.

\section{Resummation Formulas}

I will discuss only the case of single-$\mbox{\boldmath$k$}$
processes
(of DIS type) in which the
hard probe scale is denoted by $Q$ (or $2M$ if heavy-quark production
is
considered) and the
corresponding total cross-section for photon
($\mbox{\boldmath$Q$}$)-hadron (A) scattering is denoted
by $\sigma_N^{HA}(Q^2)$, where $N$ is the moment index.

Then, at high energies, the exchange of a (Regge) gluon, of
transverse
momentum
$\mbox{\boldmath$k$}$ in the electron-hadron rest frame and of
virtuality $t \simeq -
\mbox{\boldmath$k$}^2$, yields the following factorized expression
(Fig.
1a)
\beq
Q^2\sigma_N^{HA}(Q^2) = \int d^2k~~\hat\sigma^H_N
{}~(\mbox{\boldmath$k$}^2/Q^2) {\cal F}^A_N(
\mbox{\boldmath$k$})
\label{2}
\eeq
where $\hat\sigma^H_N$ is an off-shell, gauge invariant $\gamma
g(\mbox{\boldmath$k$})\rightarrow
q\bar q$ (or $Q\bar Q$) cross-section defined by the high-energy
limit
of the squared five-point
vertex in Fig. 1b, and ${\cal F}^A_N(\mbox{\boldmath$k$})$ is the
unintegrated gluon density in
hadron $A$, satisfying the BFKL integral equation$^{9)}$, and related
to
the usual density by
\beq
g^A(x,Q^2) = \int^{Q^2}_0~ d^2k~{\cal F}^A (x,\mbox{\boldmath$k$})~.
\label{3}
\eeq

Consistency of Eq. (\ref{2}) -- which is
$\mbox{\boldmath$k$}$-dependent
-- with the RG
factorization -- which is not, is demanded by requiring that the
$\mbox{\boldmath$k$}$- dependence
of ${\cal F}$ be determined by the BFKL anomalous dimension
\beq
\gamma_N(\alpha_s) \equiv \gamma \left({\bar\alpha_s\over N-1}\right)
=
{\bar\alpha_s\over N-1} + 2\zeta
(3) \left({\bar\alpha_s\over N-1}\right)^4 + \ldots~,~~~(\bar\alpha_s
=
{3\alpha_s\over\pi})~,
\label{4}
\eeq
as follows
\beq
{\cal F}^A_N(\mbox{\boldmath$k$}) = {1\over \pi
\mbox{\boldmath$k$}^2}
\gamma_N
(\alpha_s)~~\left({\mbox{\boldmath$k$}^2\over
\mu^2}\right)^{\gamma_N(\alpha_s)}~~g^A_N (\mu^2)~.
\label{5}
\eeq

\begin{figure}[H]
\hglue 1.3cm
\epsfig{figure=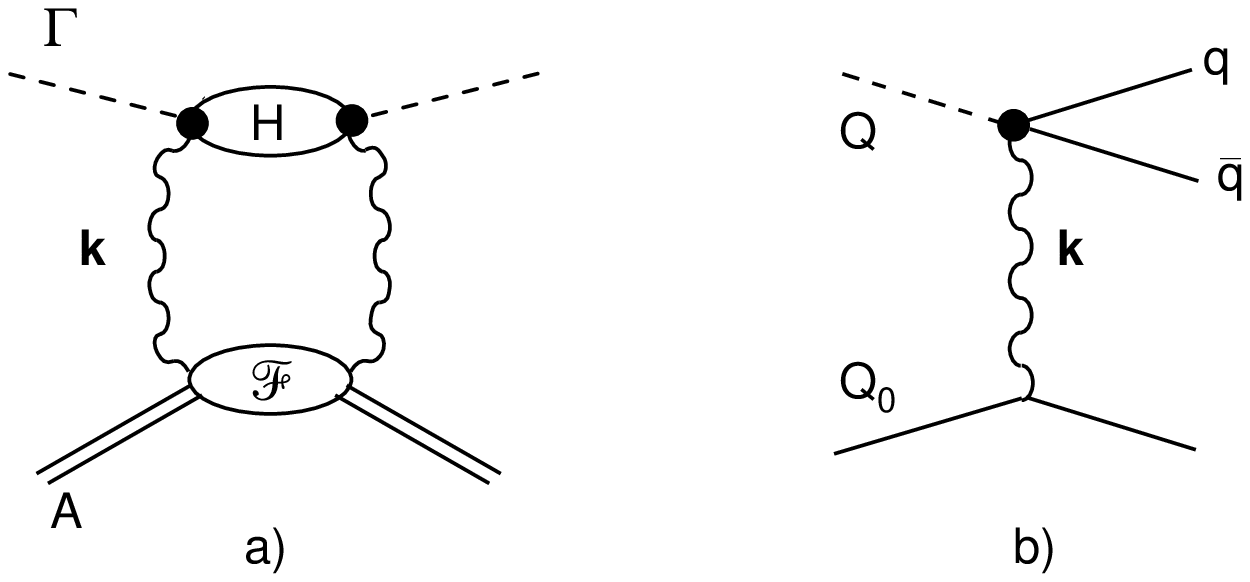,width=13cm}
\caption[]{(a) Single-$\mbox{\boldmath$k$}$ factorization diagram for
$\sigma^{H A}$ and (b)
high-energy vertex for the hard cross-section $\hat\sigma^H$. Dashed
(wavy) lines denote photon
(gluon) exchange.} \end{figure}

Inserting Eq. (\ref{5}) into Eq. (\ref{2}) allows performing the
$\mbox{\boldmath$k$}$-integration
in terms of the calculable $\mbox{\boldmath$k$}^2$-moments
\beq
h_N(\gamma ) \equiv \gamma \int^\infty_0 ~~
{d\mbox{\boldmath$k$}^2\over
\mbox{\boldmath$k$}^2}~~
\left({\mbox{\boldmath$k$}^2\over Q^2}\right)^\gamma~~
\hat\sigma^H_N\left({\mbox{\boldmath$k$}^2\over Q^2}\right)
\label{6}
\eeq
and provides the final result$^{2)}$
\beq
Q^2\sigma_N^{HA} (Q^2) = h_N
\left(\gamma\left({\bar\alpha_s(Q^2)\over
N-1}\right)\right)~
g^A_N(Q^2)~.
\label{7}
\eeq

Equation (\ref{7}) is indeed consistent with the RG, with a
coefficient
function
\beq
C_N(\alpha_s(Q^2)) = h_N \left(\gamma\left({\bar\alpha_s(Q^2)\over
N-1}\right)\right)~,
\label{8}
\eeq
which automatically resums all powers of $\alpha_s/(N-1)$ in terms of
the anomalous dimension
(\ref{5}). Thus, a lowest-order calculation of the off-shell
cross-section $\hat\sigma^H$ in Eq.
(\ref{6}) provides an all-order resummation of the coefficient
function
$C_N$ in Eq. (\ref{8})!

In deriving Eq. (\ref{7}) from Eq. (\ref{5}) we have kept, for
simplicity, $\alpha_s$ frozen and
we have used the expression (\ref{5}) even for $\mbox{\boldmath$k$}^2
<
Q^2_0$, where $Q_0$ is a
scale defining the boundary of the perturbative approach
$(\alpha_s(Q^2_0) \lappeq 1)$. It can be
proved$^{2),19)}$, however, that using a RG improved expression
(\ref{5}) and/or a full solution
of the BFKL equation$^{2),20)}$ including higher twists, and possibly
running coupling$^{21),22)}$,
{\it does~not~change} \footnote{I thus partially disagree with the
emphasis of Ref.
(20), which seems to cast doubt on this point.} the final result
(\ref{7}), except for subleading
terms of relative order $\alpha_s(Q^2)$, which are not considered
here.

Several examples of resummed formulae of type (\ref{8}) are by now
available$^{2)-5)}$. Here I
will only quote two:
\begin{itemize}
\item[(i)] Heavy flavour photoproduction$^{2)}$:
\beq
h^{Q\bar Q}(\gamma ) = {\alpha_s\over 3\pi}~~{7-5\gamma\over
3-2\gamma}~~{\Gamma (1-\gamma)^3\Gamma
(1+\gamma)\over \Gamma(2-2\gamma )}~4^{-\gamma}~,~~~(N = 0)~.
\label{9}
\eeq
\item[(ii)] DIS structure function $F_2 \equiv Q^2\sigma^{2p}$. For
scaling violations, the
following function is of particular interest$^{2),4)}$:
\bea
\gamma h_2(\gamma )\equiv \tilde h_2(\gamma )
&\equiv& \gamma \int  {d\mbox{\boldmath$k$}^2\over
\mbox{\boldmath$k$}^2}~~\left({\mbox{\boldmath$k$}^2\over
Q^2}\right)^{\gamma}~~{\partial\over \partial \log
Q^2}~~\hat\sigma^2_{N=0}~~\left(
{\mbox{\boldmath$k$}^2\over Q^2}\right)\hfill \\ \nonumber
\\ \nonumber
&= &
{\alpha_s\over 3\pi}~~{1+{3\over 2} \gamma(1-\gamma )\over 1 -
{2\over
3} \gamma}~~
{[\Gamma (1-\gamma) \Gamma (1+\gamma )]^3 \over \Gamma (2-2\gamma)
\Gamma
(2+2\gamma)}~,~~~(N=0)~.
\label{10}
\eea
\end{itemize}

Resummation effects in Eqs. (\ref{9}) and (10) can be estimated by
the
enhancement ratios \footnote{The normalization of Eq. (9) differs by
a
factor
$4^{-\gamma}$ from Ref. {2}, being referred to scale $Q^2 = 4M^2$.
The
corresponding enhancement in Eq. (11) is smaller by a factor of $2$.}
\beq
{h^{Q\bar Q}(\frac{1}{2})\over h^{Q\bar Q}(0)} = {27\over
28}~~\left({\pi\over 2}\right)^2~,\quad\quad {\tilde h_2({1\over
2})\over
\tilde h_2(0)} = {33\over 32} ~~\left({\pi\over 2}\right)^3~,
\label{11}
\eeq
in which $\gamma = 1/2$ is the asymptotic value of the anomalous
dimension
$\gamma_N$ at the BFKL Pomeron (\ref{1}). The enhancement is thus
rather
large {\it asymptotically}, ranging between 2.5 and 4, and is
still sizeable with respect to one-loop results $^{8),23)}$ in the
HERA
range.

Such important effects are due to the opening of the
$\mbox{\boldmath$k$}$ phase space (away from
the collinear region) which occurs in the high-energy regime, and is
also responsible for the
asymptotic Pomeron singularity (\ref{1}).

Is this worrying for the convergence of the perturbative series? Here
let me notice that a large
fraction of the enhancement (11) is washed out if we compare the
heavy-quark process before
with the light-quark one: indeed, their ratio is only enhanced,
asymptotically, by about 50\%.
This is because the enhancement comes from the ``disordered"
$\mbox{\boldmath$k$}$ region
$(\mbox{\boldmath$k$}^2\gg Q^2)$ mentioned before, which is
independent
of quark masses. This
remark suggests to test, experimentally, cross-section ratios, which
are
much smoother, and to
investigate, theoretically, the universality properties of the
disordered $\mbox{\boldmath$k$}$
region.

\section{Towards next-to-leading anomalous dimensions}

In order to apply the procedure above to DIS structure functions,
recall
that
the singlet anomalous dimension matrix mixes quark and gluon entries,
and
that the gluon entries
$$ {C_A\over C_F}~~\gamma_{gq} \simeq \gamma_{gg} =
\gamma_L(\bar\alpha_s (N-1))~~(1+0(\alpha_s)) $$
are leading, while the quark
ones $\gamma_{qg},\gamma_{qq}$ start at NL level, because the quark
(spin
${1\over 2}$) exchange is subleading at high energies.

Nevertheless, since the quark couples directly to the photon, while
the
gluon
coefficient carries an additional factor of $\alpha_s$, it turns out
that
quark entries are as important as gluon entries. For instance, in a
partonic
DIS-scheme, scaling violations are directly given by $\gamma_{qg}$ as
follows:
\beq
{\partial F^N_2\over \partial\log Q^2} \equiv \sum_f ~e^2_f~~\dot
q_N(Q^2) = \sum_f ~e^2_f \gamma^N_{qg}~ g_N(Q^2)~~(1+0(\alpha_s))~.
\label{12}
\eeq
Therefore, by applying $\mbox{\boldmath$k$}$-factorization to the
light-quark loop as in Eq.
(10), Catani and Hautmann$^{4)}$ found a resummed expression for
$\gamma^N_{qg}$, essentially
given by
\beq
\gamma^N_{qg} = 2N_f~\tilde
h^2_N~\left(\gamma\left({\bar{\alpha}_s\over
N-1}\right)\right)
\label{13}
\eeq
Similar considerations apply to the full GLAP equations, and yield
also
$$
\gamma^N_{qq} = {C_F\over C_A}~~\left(\gamma^N_{qg} -
{2N_f\alpha_s\over
3\pi}\right)~,
$$
thus completing the NL resummation at quark level.

On the other hand, NL terms in the gluon channel (which mix, in
principle, with the ones
considered) are much harder to obtain by
$\mbox{\boldmath$k$}$-factorization, because they involve
gluon loops and subleading gluon emission vertices, which are still
under investigation$^{24)}$.
By neglecting them, several authors have applied the resummed
formulas
(\ref{10}) and (\ref{13})
to HERA data, with encouraging results$^{8),25)}$. In the present
energy
range, they find that
resummation effects in $\gamma_{qg}$ [Eq. (\ref{13})] are more
important
than the ones in
$\gamma_{gg}$ [Eq. (\ref{4})].

There are, however, a few subtleties related to Eq. (\ref{13}). It
holds
as it stands in the
so-called $Q_0$-scheme$^{6)}$ in which the initial parton virtuality
$Q_0$ is fixed \footnote{This
can be done in the BFKL framework, because the relevant vertices are
defined in a gauge invariant
way. Extension to gluons at NL level requires some work.}, and is
used
to factorize the collinear
singularities for $Q_0 \ll Q$. In fact, if $Q_0 \not= 0$, the $\vert
\mbox{\boldmath$k$}\vert <
Q_0$ integration occurring in Eq. (\ref{2}) is automatically
suppressed.
Instead, in the $Q_0 = 0$
limit, dimensional regularization (and a MS-type scheme) is needed to
regularize and factorize all
collinear singularities. As a consequence, in the latter case Eq.
(\ref{13}) carries an additional
renormalization factor$^{19)}$ $R_N = 1+ 0(\alpha_s/(N-1))^3$, as
stated
in Ref. 4). $R_N$ departs
slowly from unity, but is soon sizeable, and is singular at the
asymptotic value $\gamma = {1\over
2}$.

The ambiguity related to the presence or absence of the $R_N$ factor
in
$\gamma_{qg}$ (or, even,
of any resummation effect at all$^{26)}$) can be regarded just as a
factorization scheme
dependence of $\gamma_{qg}$, related to a different definition of the
gluon density. Keep in mind,
however, that in any scheme, we should consider NL terms in the gluon
channel also$^{24)}$ (which
change with the scheme) and also different probes (like, e.g., $F_L,
F_3, Q\bar Q$ production, and
so on) in order to constrain the gluon density. Thus, the present
ambiguity will not last for
ever, and it is mandatory to discuss {\it all} NL terms before
reaching
firm
conclusions on resummation effects.

I personally think that the $R_N$-type factors can safely be
reabsorbed
in the initial gluon
density, because they are related to small-$x$ evolution at fixed
scale
in the leading BFKL
equation, and are thus presumably factorizable at NL level also.

More precisely, if we consider the Green's function matrix of the
BFKL
equation $G^N_{ab}(Q^2,
Q^2_0)$, $a,b = q,g$, for parton evolution from scale $Q_0$ to scale
$Q$, it differs from the
GLAP's one by a normalization matrix $K^N$, as follows$^{6)}$
\beq
G^N(Q^2,Q^2_0) = \exp\left( \int^{\log{Q^2\over
\Lambda^2}}_{\log{Q^2_0\over \Lambda^2}}
{}~\gamma^N(\alpha_s(t))\right)~~K^N(\alpha_s(Q^2_0)) + {\rm
higher~twists}~,
\label{14}
\eeq
where $K^N$ carries the information due to small-$x$ evolution around
the scale $Q_0$.

Thus we see that the high-energy properties of the BFKL equation
determine, on the one hand, the
anomalous dimensions $\gamma^N_{ab}$ providing the QCD evolution, but
also modify, on the other
hand, the initial parton densities for the GLAP equation, through the
occurrence of the matrix
$K^N$. The latter carries a hard Pomeron singularity, and possibly
unitarity corrections to
it$^{21),27)}$. Equation (\ref{14}) summarizes the leading twist
consistency properties of the
BFKL equation with the renormalization group.
\vfill\eject
\noindent
{\bf Acknowledgements}

It is a pleasure to thank Stefano Catani and Francesco Hautmann for
valuable comments and the
organizers of the Blois Conference for the warm atmosphere enjoyed at
the meeting. This work was
supported in part by E.C. ``Human Capital and Mobility" contract \#
ERBCHRXCT 930357.

\vspace*{2cm}
\noindent
{\bf Bibliography}
\begin{itemize}
\item[~1)] ZEUS Collaboration, M. Derrick et al., \ZP {\bf C65}
(1995)
379;\\
H1 Collaboration, T. Ahmed et al., \NP {\bf B439} (1995) 471.
\item[~2)] S. Catani, M. Ciafaloni and F. Hautmann, \PL {\bf B242}
(1990) 97; \NP {\bf B366}
(1991) 135.
\item[~3)] J.C. Collins and R.K. Ellis, \NP {\bf B360} (1991) 3.
\item[~4)] S. Catani and F. Hautmann, \PL {\bf B315} (1993) 157; \NP
{\bf B427} (1994) 475.
\item[~5)] G. Camici and M. Ciafaloni, \NP {\bf B420} (1994) 615.
\item[~6)] M. Ciafaloni, CERN Preprint CERN-TH.95-119, to be
published
in \PL B.
\item[~7)] P. Nason, S. Dawson and R.K. Ellis, \NP {\bf B303} (1988)
607.
\item[~8)] R.K. Ellis, F. Hautmann and B.R. Webber, Preprint
Cavendish-HEP-94/18.
\item[~9)] L.N. Lipatov, \SJNP {\bf 23} (1976) 338;\\
E.A. Kuraev, L.N. Lipatov and V.S. Fadin, \SP JETP {\bf 45} (1977)
199;\\
Ya. Balitskij and L.N. Lipatov, \SJNP {\bf 28} (1978) 822.
\item[10)] H. Cheng and T.T. Wu, Expanding Protons: Scattering at
High
Energies (MIT Press,
Cambridge, MA) and references therein.
\item[11)] V.N. Gribov and V.N. Lipatov, \SJNP {\bf 15} (1972) 438,
675;\\
G. Altarelli and G. Parisi, \NP {\bf B126} (1977) 298;\\
Yu.L. Dokshitzer, \SP JETP {\bf 46} (1977) 641.
\item[12)] See, e.g., A.J. Askew, J. Kwiecinski, A.D. Martin and P.J.
Sutton, \PR {\bf D47}
(1993) 3775; {\bf D49} (1994) 4402.
\item[13)] T. Jaroszewiecz, \APP {\bf B11} (1980) 965.
\item[14)] See, e.g., A. Bassetto, M. Ciafaloni and G. Marchesini,
{\it
Physics Reports} {\bf
100} (1983) 201 and references therein.
\item[15)] M. Ciafaloni, \NP {\bf B296} (1987) 249.
\item[16)] L.V. Gribov, Yu.L. Dokshitzer, S.I. Troyan and V.A. Khoze,
\SP JETP {\bf 67} (1988)
1303.
\item[17)] S. Catani, F. Fiorani and G. Marchesini, \PL {\bf 234B}
(1990) 339; \NP {\bf B336}
(1990) 18.
\item[18)] J. Kwiecinski, A.D. Martin and P.J. Sutton, Durham
Preprint
DTP/95/22.
\item[19)] S. Catani, M. Ciafaloni and F. Hautmann, \PL {\bf B307}
(1993) 147.
\item[20)] J. Kwiecinski and A.D. Martin, Durham Preprint DTP/95/34.
\item[21)] L.V. Gribov, E.M. Levin and M.G. Ryskin, {\it Physics
Reports} {\bf 100} (1983) 1.
\item[22)] J. Kwiecinski, \ZP {\bf C29} (1985) 561;\\
J.C. Collins and J. Kwiecinski, \NP {\bf B316} (1989) 307.
\item[23)] S. Catani, M. Ciafaloni and F. Hautmann, Proceedings HERA
Workshop (DESY, Hamburg,
1991), p. 690.
\item[24)] V.S. Fadin and L.N. Lipatov, \YF {\bf 50} (1989) 1141; \NP
{\bf B406} (1993) 259;\\
V.S. Fadin and R. Fiore, \PL {\bf B294} (1992) 286.
\item[25)] R. Ball and S. Forte, CERN Preprints CERN-TH/95-01 and
CERN-TH/95-148 (to be published
in \PL B).
\item[26)] S. Catani, Firenze Preprint DFF-226/5/95.
\item[27)] J. Bartels, \PL {\bf B298} (1993) 204;\\
L.N. Lipatov, Padova Preprint DFPD-93-TH-70 (1993), JETP {\it Lett.}
{\bf 59} (1994) 596;\\
L.D. Faddeev and G.P. Korchemsky, Stony Brook Preprint ITP-SP-94-14;
\PL
{\bf B342} (1995) 311.
\end{itemize}

\end{document}